\documentclass[acmsmall]{acmart}

\usepackage{array}
\usepackage{appendix}
\usepackage{booktabs} 
\usepackage{xcolor} 
\usepackage{soul} 

\AtBeginDocument{%
  \providecommand\BibTeX{{%
    \normalfont B\kern-0.5em{\scshape i\kern-0.25em b}\kern-0.8em\TeX}}}

\begin{document}

\title{Where Responsible AI meets Reality: Practitioner Perspectives on Enablers for Shifting Organizational Practices}

\author{Bogdana Rakova}
\affiliation{
  \institution{Partnership on AI}
  \streetaddress{115 Sansome St 1200}
  \city{San Francisco}
  \country{USA}
}
\affiliation{
  \institution{Accenture}
  \streetaddress{415 Mission St Floor 35}
  \city{San Francisco}
  \country{USA}}
\email{b.rakova@gmail.com}

\author{Jingying Yang}
\affiliation{
  \institution{Partnership on AI}
  \streetaddress{115 Sansome St 1200}
  \city{San Francisco}
  \country{USA}}
\email{jingying@partnershiponai.org}

\author{Henriette Cramer}
\affiliation{
  \institution{Spotify}
  \streetaddress{4 World Trade Center}
  \city{New York, NY 10007}
  \country{USA}}
\email{henriette@spotify.com}

\author{Rumman Chowdhury}
\affiliation{
  \institution{Accenture}
  \streetaddress{415 Mission St Floor 35}
  \city{San Francisco}
  \country{USA}}
\email{rumman.chowdhury@accenture.com}

\renewcommand{\shortauthors}{Bogdana Rakova et al.}

\setcopyright{acmcopyright}
\acmJournal{PACMHCI}
\acmYear{2021} \acmVolume{5} \acmNumber{CSCW1} \acmArticle{7} \acmMonth{4} \acmPrice{15.00}\acmDOI{10.1145/3449081}

\begin{abstract}
Large and ever-evolving technology companies continue to invest more time and resources to incorporate responsible Artificial Intelligence (AI) into production-ready systems to increase algorithmic accountability. This paper examines and seeks to offer a framework for analyzing how organizational culture and structure impact the effectiveness of responsible AI initiatives in practice. We present the results of semi-structured qualitative interviews with practitioners working in industry, investigating common challenges, ethical tensions, and effective enablers for responsible AI initiatives. Focusing on major companies developing or utilizing AI, we have mapped what organizational structures currently support or hinder responsible AI initiatives, what aspirational future processes and structures would best enable effective initiatives, and what key elements comprise the transition from current work practices to the aspirational future. 
\end{abstract}


\maketitle

\section{Introduction}
While the academic discussion of algorithmic bias has an over 20-year long history ~\cite{friedman1996bias}, we have now reached a transitional phase in which this debate has taken a practical turn. The growing awareness of algorithmic bias and the need to responsibly build and deploy Artificial Intelligence (AI) have led increasing numbers of practitioners to focus their work and careers on translating these calls to action within their domains~\cite{madaio20, holstein2019improving}. New AI and machine learning (ML) responsibility or fairness roles and teams are being announced, product and API interventions are being presented, and the first public successes \textemdash and lessons learned \textemdash are being disseminated ~\cite{haydn2020}. However, practitioners still face considerable challenges in attempting to turn theoretical understanding of potential inequities into concrete action ~\cite{holstein2019improving, Krafft20}. 

Gaps exist between what academic research prioritizes, and what practitioners need. The latter includes developing organizational tactics and stakeholder management~\cite{holstein2019improving, tutorialfat} rather than technical methods alone. Beyond the need for domain-specific translation, methods, and technical tools, responsible AI initiatives also require operationalization within \textemdash or around \textemdash existing corporate structures and organizational change. Industry professionals, who are increasingly tasked with developing accountable and responsible AI processes, need to grapple with inherent dualities in their role~\cite{metcalfowners} as both agents for change based on their own values and/or their official role, but also as workers with careers in an organization with potentially misaligned incentives that may not reward or welcome change ~\cite{madaio20}. Most commonly, practitioners have to navigate the interplay of their organizational structures and algorithmic responsibility efforts with relatively little guidance. As Orlikowski points out, whether designing, appropriating, modifying, or even resisting technology, human agents are influenced by the properties of their organizational context ~\cite{orlikowski1992duality}. This also means that some organizations can be differentially successful at implementing organizational changes. Individuals' strategies must adapt to the organizational context and follow what is seen as successful and effective behavior within that setting. Myerson for example describes the concept of "tempered radicals." These are employees who slowly but surely create corporate change by pushing organizations through persistent small steps. Advocating for socially responsible business practices became part of tempered radicals' role over time. These employees create both individual and collective action, relying on their own perceived legitimacy, influence and support built within their organizational context ~\cite{meyerson2004tempered}.

Interestingly, the tension between academic research and industry practice is visible in research communities such as FAccT, AIES, and CSCW, where people answering calls to action with practical methods are sometimes met with explicit discomfort or disapproval from practitioners working within large corporate contexts. Vice versa,  practitioners affecting concrete change in practice may have achieved such results in ways that may not fit external research community expectations or norms. Within the discourse on unintended consequences of ML-driven system, we have seen both successes and very public failures \textemdash even within the same corporation ~\cite{haydn2020} \textemdash making it imperative to understand such dynamics.

 This paper builds on the prior literature in both organizational change and algorithmic responsibility in practice to better understand how these still relatively early efforts are taking shape within organizations. We know that attention to the potential negative impacts of machine learning is growing within organizations, but how to leverage this growing attention to effectively drive change in the AI industry remains an open question. To this end, we present a study involving 26 semi-structured interviews with professionals in roles that involve concrete projects related to investigating responsible AI concerns or "fair-ML" (fairness-aware machine learning ~\cite{selbst2019fairness}) in practice. We intend this to refer not only to fairness-related projects but also more broadly projects related to the work on responsible AI and accountability of ML products and services given the high degree of overlap in goals, research, and people working on these topics.

Using the data from the semi-structured qualitative interviews to compare across organizations, we describe prevalent, emergent, and aspirational future states of organizational structure and practices in the responsible AI field, based on how often respondents identified the practice during the interview and whether the practice is currently existing or is a desired future change. We investigate practitioners' perceptions of their own role, the role of the organizational structures in their context, and how those structures interact with adopting responsible AI practices. Based on those answers, we identify four major questions that organizations must now adapt to answer as responsible AI initiatives scale. Furthermore, we describe how respondents perceived transitions occurring within their current contexts, focusing on organizational barriers and enablers for change. Finally, we present the outcome of a workshop where attendees reflected upon early insights of this study through a structured design activity.

The main contribution of our work is the qualitative analysis of semi-structured interviews about the responsible AI work practices of practitioners in industry. We found that most commonly, practitioners have to grapple with lack of accountability, ill-informed performance trade-offs and misalignment of incentives within decision-making structures that are only reactive to external pressure. Emerging practices that are not yet widespread include the use of organization-level frameworks and metrics, structural support, and proactive evaluation and mitigation of issues as they arise. For the future, interviewees aspired to have organizations invest in anticipating and avoiding harms from their products, redefine results to include societal impact, integrate responsible AI practices throughout all parts of the organization, and align decision-making at all levels with an organization's mission and values. Preliminary findings were shared at an interactive workshop during a large machine learning conference, which yielded organizational level recommendations to (1) create veto ability across levels, (2) coordinate internal and external pressures, (3) build robust communication channels between and within levels of an organization, and (4) design initiatives that account for the interdependent nature of the responsible AI work practices we have heretofore discussed.

\section{Literature review}

\subsection{Algorithmic responsibility in practice}
An almost overwhelming collection of principles and guidelines have been published to address the ethics and potential negative impact of machine learning. Mittelstadt et al.~\cite{mittelstadt2019ai} discuss over sixty sets of ethical guidelines, Zeng et al. \cite{zeng2018linking} provide a taxonomy of 74 sets of principles, while Jobin et al. find 84 different sets of principles \cite{jobin2019global}. Even if there is relative, high-level agreement between most of these abstract guidelines ~\cite{jobin2019global, zeng2018linking}, how they are translated into practice in each context remains very unclear ~\cite{mittelstadt2019ai}. Insight is available from how companies changed their practices in domains such as privacy and compliance in response to legislative directives ~\cite{privacybook}. The active debate on how requirements in the EU's GDPR are to be interpreted ~\cite{kaminski2020multi, malgieri2020concept}, however, illustrate the challenges of turning yet nascent external guidance into concrete requirements. Krafft et al. ~\cite{Krafft20} point out that even between experts, there is a disconnect between policymakers and researchers' definitions of such foundational terms as `AI'. This makes the application of abstract guidelines even more challenging and raises the concern that focus may be put on future, similarly abstract technologies rather than current, already pressing problems.

The diverse breadth of application domains for machine learning suggests that requirements for applying guidelines in practice should be steered by the specific elements of the technologies used, specific usage contexts, and relevant local norms  ~\cite{mittelstadt2019ai}. Practitioners encounter a host of challenges when trying to perform such work in practice ~\cite{holstein2019improving}. Organizing and getting stakeholders on board are necessary to be able to drive change ~\cite{tutorialfat}. This includes dealing with imperfection, and realizing that tensions and dilemmas may occur when "doing the right thing" does not have an obvious and widely agreed upon answer ~\cite{Fazelpour20, cramer2018assessing}. It can be hard to foresee all potential consequences of systems while building them, and it can be equally difficult to identify how to overcome unwanted side-effects, or even why they occur technically ~\cite{googleAudit2020}. A fundamental challenge is that such assessment should not simply be about technical, statistical disparities, but rather active engagement to overcome the lack of guidance decision-makers have on what constitute "just" outcomes in non-ideal practice ~\cite{Fazelpour20}. Additional challenges include organizational pressures for growth, common software development approaches such as agile working that focus on rapid releases of minimal viable products, and incentives that motivate a focus on revenue within corporate environments ~\cite{holstein2019improving,madaio20, haydn2020}. Taking inspiration from other industries where auditing processes are standard practice still means that auditing procedures have to be adjusted to product and organizational contexts, and require defining the goal of the audit in context ~\cite{googleAudit2020}. This means that wider organizational change is necessary to translate calls to action into actual process and decision-making.

\subsection{Organizational change and internal/external dynamics}
Current challenges faced by responsible AI efforts can be compared to a wide selection of related findings in domains such as legal compliance~\cite{trevino1999managing} where questions arise regarding whether compliance processes actually lead to more ethical behavior~\cite{krawiec2003cosmetic}, diversity and inclusion in corporate environments ~\cite{barak2016managing, kalev2006best}, and corporate privacy practices ~\cite{privacybook}. All of these domains appear to have gone through a process that is mirrored in current algorithmic responsibility discussions: publication of high-level principles and values by a variety of actors, the creation of dedicated roles within organizations, and urgent questions about overcoming challenges and achieving "actual" results in practice and how to avoid investing in processes that are costly but do not deliver beyond cosmetic impact.

 As Weaver et al. pointed out in 1999 ~\cite{weaver1999corporate}, in an analysis of the Fortune 1000 ethics practices, success relies not only on centralized principles, but also their diffusion into managerial practices in the wider organization. Interestingly, while external efforts can effectively put reputational and legislative pressure on companies, internal processes and audits are just as important, and they all interact. Internally, this is apparent in the process of legitimization of the work on `tempered radicals' in Myerson's work ~\cite{meyerson2004tempered} as described in the introduction, and these radicals' internal journey. External forces can help in more or less productive ways in that process.
 As discussed by Bamberger and Mulligan ~\cite{privacybook}, for corporate privacy efforts in particular, both external and internal forces are necessary for work on corporate responsibility to be effective. Internally, they suggest focusing on getting onto board-level agendas to ensure attention and resourcing, having a specific boundary-spanning privacy professional to lead adoption of work practices, and ensuring `managerialization' of privacy practices by increasing expertise within business units and integration within existing practices. Externally, they suggest that creating positive ambiguity by keeping legislation broad can push more accountability onto firms for their specific domains, which can create communities and promote sharing around privacy failures. They found that ambiguity in external privacy discussions could foster reliance on internal professionals' judgements, and thus created autonomy and power for those professionals identified as leading in privacy protection. Thus, they illustrate how ambiguity \textemdash rather than a fully defined list of requirements \textemdash can actually help promote more reflection and ensure that efforts go beyond compliance.

A similar internal/external dynamic is visible within the algorithmic responsibility community. For example, in the Gender Shades project, Buolamwini and Gebru~\cite{buolamwini2018gender} presented not only an external audit of facial recognition APIs, but also reactions from the companies whose services were audited to illustrate more and less effective responses. Such external audits can result in momentum inside of companies to respond to external critique, and in selected cases, to make concrete changes to their products. Internal efforts in turn have access to more data, ensure that auditing can be completed before public releases, develop processes for companies, and allow companies to take responsibility for their impact ~\cite{googleAudit2020}. Successes are beginning to emerge, and have ranged from positive changes on policy and process resulting from corporate activism, tooling built for clients or internal purposes, to direct product "fixes" in response to external critique ~\cite{haydn2020}. For example, Raji et al.~\cite{googleAudit2020} present an extensive algorithmic auditing framework developed by a small team within the larger corporate context of Google. They offer general methods such as data and model documentation ~\cite{modelcards} and also tools such as metrics to enable auditing in specific contexts like image search ~\cite{Mitchell20metrics}. Implementing these methods and tools then requires corporate processes to provide the resources for such auditing and to ensure that results of audits impact decisions within the larger organizational structure.

\subsubsection{Organizational research and structures}
To situate our work in this broader context, we will briefly examine different perspectives on organizational structures. First, it is worthwhile to revisit what organizational theorist Wanda Orlikowski~\cite{orlikowski1992duality} called the duality of technology in organizations. Orlikowski discusses how people in organizations create and recreate meaning, power and norms. Orlikowski's `structurational' model of technology comprises of these human agents, the technology that mediates their task execution, and the properties of organizations. The latter institutional properties range from internal business strategies, control mechanisms, ideology, culture, division of labor and procedures, communication patterns, as well as outside pressures such as governmental regulation, competition and professional norms, and wider socio-economic conditions. People's actions are then enabled and constrained by these structures, which are themselves the product of previous human actions. This perspective was augmented by Orlikowski  ~\cite{Orlikowski2000} to include a practice orientation; repeated interactions with technologies within the specific circumstances also enact and form structures. 

Similarly, Dawson provides an extensive review of perspectives in studies on organizational change ~\cite{dawson2019reshaping} and discusses the `process turn', where organizations are seen as ever-changing rather than in discrete states; what may appear as stable routines may in actuality be fluid. Dawson emphasizes the socially constructed process, and the subjective lived experiences: actors' collaborative efforts in organizations unfold over time and dialogue between them shapes interpretations of changes. 
Such dynamics are also present in what organizational theorist Richard Scott \cite{scott2015organizations} summarized as the rational, natural, and open perspective on organizations. `Rational' organizations were seen as `the machine', best suited to industries such as assembly line manufacturing where tasks are specified by pre-designed workflow processes. The `natural' organization signified a shift in organizational ideology. No longer were people seen as mere appendages to the machines, but rather as crucial learners in relationship with machines. The metaphor is that of the organization as an 'organism' with a strong interior vs. exterior boundary, and needs to `survive'. Similar to an organism, the organisation grows, learns, and develops. As a consequence of the survival ideology, the exterior environment can be seen as a threat against which the organism must adapt to survive. Scott however describes how the notion of `environment as threat' was replaced by the realization that environmental features are the conditions for survival. The central insight emerging from `open' systems thinking is that all organizations are incomplete and depend on exchanges with other systems. The metaphor became that of an `ecology'. Open systems are characterized by (1) interdependent flows of information and (2) interdependent activities, performed by (3) a shifting coalition of participants by way of (4) linking actors, resources and institutions, in order to (5) solve problems in (6) complex environments. For responsible AI efforts to succeed then, organizations must successfully navigate the changes necessary within `open' systems.

\subsubsection{Multi-stakeholder communities as meta organizational structures}

The described `ecologies', particularly in `open' systems, contain formal and informal meta-organizational structures, which have been studied in other contexts and are of increasing growing importance to the field of responsible AI. Organizations often interact with each other through standards bodies, communities, processes, and partnerships. These meta-processes can have as goals (1) producing responses to proposed regulations, standards, and best practices, (2) fostering idea exchange between silos, and (3) self-regulation. Organizations participate in multi-stakeholder initiatives to achieve a number of their own goals, including advocating for business interests, keeping up to date on industry trends, and having a voice in shaping standards or regulations that they will then be subjected to. 

Berkowitz ~\cite{Berkowitz2018} discusses the shift towards governance in sustainability contexts, and the key role that meta-organizations can have in facilitating meta-governance of corporate responsibility beyond simply complying with legislation. She identifies six capabilities needed for sustainable innovations: (1) anticipation of changes and negative impacts of innovation, (2) resilience to changes, (3) reflexivity, (4) responsiveness to external pressures and changing circumstances, (5) inclusion of stakeholders beyond immediate decision makers, and (6) comprehensive accountability mechanisms. Meta-organizations can promote inter-organizational learning and building of these six capabilities.

Similarly, within the field of AI, multi-stakeholder organizations, standards, and self-organized projects have been created in recent years to acknowledge the need for interdisciplinary expertise to grapple with the wide reaching impacts of AI on people. Many AI researchers have been vocal proponents of expanding the number of perspectives consulted and represented, including stakeholders such as policymakers, civil society, academics from other departments, impacted users, and impacted nonusers. Reconciling perspectives from diverse stakeholders presents its own set of challenges that change depending on the structure of the organization. Participatory action offers relevant frameworks for characterizing options for decision making in multistakeholder contexts. Decision making can be centralized within a formal organization with stakeholders being informed, consulted, involved, collaboration, or else stakeholders can self-organize informally to achieve the same levels of participation. The structures present at a meta organizational level will differ and enable the application of different group-level decision making processes. For example, ad hoc groups of researchers have self-organized to create unconference events and write multi-stakeholder reports, including reports with large groups of authors (e.g. ~\cite{brundage2020trustworthy}) based originally on discussions within workshops held under Chatham house rules, while others have created new formal organizations, conferences such as AIES or FAccT, or research institutes.

In a similar manner to Berkowitz ~\cite{Berkowitz2018}, we focus here on the "how" of achieving more adoption of responsible AI work practices in industry. We further investigate how practitioners experience these changes within the context of different organizational structures, and what they see as the shifts that drive or hinder their work within their organizations.


\section{Study and Methods} \label{studyandmethods} 

Our motivation for this work was to identify enablers that could shift organizational change towards adopting responsible AI practices. Responsible AI research has influenced organizational practices in recent years, with individuals and groups within companies increasingly tasked with implementing research into action, whether formally or informally. Our research applies theories and frameworks of organizational structure and change management to characterize the growing practice of applied responsible AI. To better understand the implications of organizational structure on day-to-day responsible AI work and outcomes, we interviewed practitioners who are actively involved in these initiatives by themselves or within a larger team. 

We conducted 26 semi-structured interviews with people based in 4 continents from 19 organizations. Except for two 30 minute interviews, all other interviews lasted between 60 and 90 minutes. Participants were given a choice of whether to allow researchers to record the interview for note taking purposes. A total of 11 interviews were recorded. In cases where the interview was not recorded, we relied on writing down the respondents' answers to the questions during the course of the interview. In several cases, participants requested to additionally validate any written notes and make necessary clarifications before their use in the study to ensure that their anonymity was not compromised. 

\begin{table}[]
\caption{Distribution of interviewee job functions and keywords interviewees used to describe their roles.}
\label{tab:roles-table}
\begin{tabular}{ p{3cm}p{2cm}p{8cm}  }

\toprule
Role                      & Respondents                   & Responsible AI / fair-ML workstream framing                                                                                                                                                                                                                                             \\ \midrule
AI Strategy               & R16, R25            & thought leadership; strategic planning; building external relationships; working with operating models; proactively figuring out the pain points and creating solutions;                                                                                                                                                                                   \\
Engineering               & R1, R14,  R19, R21            & fairness evaluations; internal checklists; implementing new capabilities; data science;                                                                                                                                                                                   \\
Human Resources           & R12, R13                      & assessment innovation; talent innovation research;   \\
Legal                     & R8, R20, R26                   & policy; legal counseling; investigating legal issues and questions; responsible AI; privacy; ethical and governance guidelines; comprehensive pillars; digital ethics;                                                                                                                                                                                                                 \\
Marketing and Sales       & R10, R24                      & algorithmic accountability; understand and explain what an algorithm does; fairness auditing and explainability in terms of bias;                                                                                                                                      \\
ML Research & R17, R22, R23                 & algorithmic audits; explainability; social impact of AI; sociotechnical systems; educational efforts; fairness;                                                 \\
Policy                    & R4, R5, R6, R18               & distribution of benefits from AI; norms; communication ability and navigating external expecations; fairness; mitigation of risk;  \\
Product Management        & R2, R3, R7, R9, R11, R15 & reconsidering the ML lifecycle; interpretability; influenced by broader industry trends; practical needs; responsible AI; ethics of technology; ethics review; auditing; bias assessment;                                                                                                                                                                                                                                             \\ \bottomrule
\end{tabular}
\end{table}

\subsubsection{Sampling technique}
Participants were recruited through a convenience sampling technique through snowball sampling from participants who recommended other interviewees. Three recruiting criteria were used to find interviewees: (1) did they work closely with product, policy, and/or legal teams, (2) did the outputs of their work have a direct impact on ML products and services, and (3) were some aspects of their work related to the field of responsible AI. We filtered out individuals whose roles were solely research, although interviewees may also be active contributors to responsible AI research in addition to their existing work stream.

Through the ongoing conversations we had with practitioners before as well as after conducting the qualitative interviews, we aimed to establish a substantial level of trust and transparency, which we felt was necessary given the sensitive nature of the topics discussed. This allowed for more open, nuanced, and in-depth discussions where practitioners felt that there is a shared understanding between interviewers and interviewees about the often unvoiced challenges in responsible AI work.

We intentionally sought to interview as diverse a group of practitioners as possible to capture perspectives from a broad range of organizational contexts. In \emph{Table~\ref{tab:roles-table}} we summarize the functional roles of the interviewees who participated in the project and how they describe their responsible AI work. Participants came from a wide variety of functions, including AI Strategy, Engineering, Human Resources, Legal, Marketing and Sales, Machine Learning Research, Policy, and Product Management. Among the 26 participants, ten had educational background in Social Science, eight in Computer Science, seven in Law and Policy, and one practitioner had a degree in Economics. The majority of respondents were geographically located in the US (21 out of 26), two participants were in the UK, and the rest of the respondents were based in Australia, Denmark, and Japan. The average length that interviewees have been with their organization is 5 years and 5 months, where more than one third of the practitioners have been with their company for more than five years (9 people) and 2 people spent decades with their organization. Lastly in terms of organizational sectors, 11 practitioners worked in business-to-business organizations, 2 in business-to-consumer, and 13 in organizations which were both business-to-business and business-to-consumer.

\subsubsection{Interview protocol}
The script and questions for the semi-structured interviews were reviewed by an industrial-organizational psychologist and responsible AI practitioners within three different organizations. Questions were grouped into different sections, exploring the current state of responsible AI work, the evolution of the work over time, how the work is situated within the organization, how responsibility and accountability for the work are distributed, performance review processes and incentives, and what desired aspirational future structures and processes would enable more effective work. The semi-structured nature of the interview provided standard questions that were asked of all participants while allowing interviewers the flexibility to follow up on interesting insights as they arose during interviews. The full set of questions can be found in \emph{Appendix \ref{questionnaire}}. 

\subsubsection{Analysis}
To analyze the interview data, we utilized a standard methodology from contextual design - interpretation session and affinity diagramming \cite{holtzblatt1997contextual}. Through a bottom up affinity diagramming approach, we iteratively assigned codes to various concepts and themes shared by the interviewees. We iteratively grouped these codes into successive higher level themes and studied the relationships between them.

\subsubsection{Workshop}
In addition to semi-structured interviews, we organized a workshop at a machine learning conference attended by a highly self-selected group of people interested in responsible AI from industry, academia, government, and civil society. The first half of the workshop was a presentation of preliminary insights from the literature review and results sections of this paper. We then conducted an interactive design exercise where participants were organized into 13 groups of between 4 and 6 individuals per group. Each group was given a scenario description of an AI organization that exemplified the prevalent work practices discussed in the \emph{Results: Interviews} section below. The facilitators guided groups through a whiteboard discussion of the following questions:
\begin{itemize}
  \item What are examples of emerging responsible AI work practices in the context of the scenario?
  \item What are examples of structures or processes in the prevalent organizational structure which are outside of the scope of responsible AI work but which act to protect and enable emerging fair-ML practices?
  \item What are examples of outlier practices outside of the prevalent practices in the scenario?
  \item What connections exist between these practices and organizational structures?
  \item What practices or organizational structures could enable positive self-reinforcing outcomes through making the connections stronger?
\end{itemize}
The workshop activity was designed to allow participants to (1) gain a deeper understanding of the responsible AI challenges by connecting study findings to their own experiences, (2) collaboratively explore what organizational structures could enable the hypothetical organization developing AI products and services to resolve them through, and (3) map interdependencies and feedback loops that exist between practices to identify potentially effective recommendations to address the challenges of implementing responsible AI initiatives.

\section{Results: Interviews} \label{results}
We start with a high level overview of our findings followed by a discussion of the key themes that emerged from the conducted interviews. 

\subsection{Overview}
About a quarter of the participants had initiated their responsible AI work in their current organization within the past year (7 out of 26) compared to 73\% (19 out of 26) worked on efforts that had started more than an year ago. More than half of the interviewees worked on their initiatives as individuals and not as part of a team (14 out of 26). About 40\% of the respondents reported that they  volunteer time outside of their official job function to do their work on responsible AI initiatives (11 out of 26) while the remaining 15 out of 26 participants had official roles related to responsible AI. Among the 15 interview participants with official roles related to responsible AI, 8 individuals were externally hired into their current role, while 7 transitioned into it from other roles within their organization. Interviewees who changed the focus of their existing roles or transitioned into responsible AI-related role were most commonly previously in project management roles (4 out of 7), then research (2 out of 7), then legal (1 out of 7). The majority of participants who had official responsible AI-related roles reported benefiting from an organizational structure that allowed them to craft their own role in a very dynamic and context-specific way.

Since the beginning of our conversations, we noticed that practitioners used different language in the way they described their work and how it relates to responsible AI. We observed commonalities in the way practitioners from each function framed their responsible AI work (see \emph{Table~1}). For example, while project managers described their work in terms of product life-cycles and industry trends, legal practitioners discussed the responsible AI aspects of their role in terms of comprehensive pillars and ethical governance guidelines.

We note that a few interviewees described going through stress-related challenges in relation to their responsible AI work. During some of the interviews, we saw a noticeable tone change in the interviewees' voice when discussing questions related to ethical tensions, accountability, risk culture, and others. Furthermore, some respondents have left their organizations between when we conducted the interviews in late 2019 and when we submitted this paper in October 2020. While we acknowledge the nascent state of responsible AI functions, these observations could point to opportunities for further study.

There were various common perspectives that we heard practitioners express repeatedly. We saw the need for a multi-faceted thematic analysis which encompasses three intuitive clusters of data: (1) currently dominant or prevalent practices, (2) emerging practices, and (3) aspirational future context for responsible AI work practices in industry:

\begin{itemize}
\item The prevalent practices comprise what we saw most commonly in the data.
\item The set of emerging practices includes practices which are shared among practitioners but less common than prevalent practices.
\item The aspirational future consists of the ideas and perspectives practitioners shared when explicitly asked about what they envision for the ideal future state of their work within their organizational context.
\end{itemize}

Within the thematic analysis (see \emph{Table~\ref{table:1_overview}}), we found four related but distinct key questions that every organization must have processes and structures to support answering: 
\begin{itemize}
    \item When and how do we act?
    \item How do we measure success?
    \item What are the internal structures we rely on?
    \item How do we resolve tensions?
\end{itemize}
As organizations seek to scale responsible AI practices, they will have to transition from the prevalent or emerging practices of answering these questions to the structures and processes of the aspirational future. It is important to note that not all emerging practices we found in the data will necessarily lead to the aspirational future. In what follows, we provide details about practitioners' personal perspectives and experiences within the individual themes and questions.

\begin{table}[]
\caption{Trends in the common perspectives shared by diverse responsible AI practitioners.}
\label{tab:overview-table}
\begin{tabular}{ p{2cm}p{3cm}p{3cm}p{4cm}  }
 \hline
 \setlength{\tabcolsep}{20pt}
  & Prevalent practices & Emerging practices & Aspirational future \\ [0.5ex] 
 \hline\hline
 \setlength{\tabcolsep}{20pt}
When and how do we act? 
 & \textbf{Reactive} \newline
Organizations act only when pushed by external forces (e.g. media, regulatory pressure)
 & \textbf{Proactive} \newline
Organizations act proactively to address potential responsible AI issues
 & \textbf{Anticipatory} \newline
Organizations have deployed frameworks that allow for anticipating risks 
 \\ 
 \hline
How do we measure success? 
& \textbf{Performance \newline trade-offs} \newline 
Org-level conversations about responsible AI dominated by ill-informed performance trade-offs
& \textbf{Provenance} \newline 
Org-level metrics frameworks and processes are implemented to evaluate responsible AI projects
& \textbf{Concrete results} \newline
Concepts of results are redefined to include societal impact through data-informed efforts
\\
 \hline
What are the internal structures we rely on? 
& \textbf{Lack of \newline accountability} \newline
Responsible AI work falls through the cracks due to role uncertainty
& \textbf{Structural support} \newline
Scaffolding to support responsible AI work begins to be erected on top of existing internal structures
& \textbf{Integrated} \newline 
Responsible AI work is integrated throughout all business processes related to product teams
\\
 \hline
How do we resolve tensions? 
& \textbf{Fragmented} \newline 
Misalignment between individual and team incentives and org-level mission statements
& \textbf{Rigid} \newline
Overly rigid organizational incentives demotivate addressing ethical tensions in responsible AI work
& \textbf{Aligned} \newline
Ethical tensions in work are resolved in accordance with org-level mission and values \\ [1ex]
 \hline
\end{tabular}
\label{table:1_overview}
\end{table}

\subsection{When and how do we act?}
One transition we identified in the data is how organizations choose when and how to act. This includes questions of who chooses to prioritize what information within which decision-making processes. We found that many organizations behave reactively, fewer are now proactive, and respondents aspire for their organizations to become anticipatory in the future.
\subsubsection{Prevalent work practices}
Most commonly, interviewees described responsible AI work in their organizations as reactive. The most prevalent incentives for action were catastrophic media attention and decreasing media tolerance for the status quo. Many participants reported that responsible AI work can be perceived as a "taboo topic" in their organizations. Raising awareness was a challenge for one interviewee who shared: "It was an organizational challenge for us, it's hard as when something is so new - we run into 'Whose job is this?'" when they bring up topics about algorithmic fairness or inequity in harm at work. We found that the uncertainty and unwillingness to engage in a deeper understanding of responsible AI issues may lead to unproductive discussions or outright dismissal of important but often unvoiced concerns. responsible AI work is often not compensated, as in the case of the 40\% of respondents volunteering their time to work on responsible AI initiatives, or is perceived as ambiguous or too complicated for the organization's current level of resources. In response to the question about how interviewees are recognized for their work, one interviewee shared: "many of the people volunteering with our team had trouble figuring out how to put this work in the context of their performance evaluation." In several cases, the formation of a full-time team to conduct responsible AI work was only catalyzed by the results from \textit{volunteer}-led investigations of potential bias issues within models that were en route to deployment. The volunteers for these investigations went far beyond their existing role descriptions, sometimes risking their own career progression, to take on additional uncompensated labor to prevent negative outcomes for the company. This highlights the reactive nature of organizational support for responsible AI work in prevalent practice. Legal compliance was another factor that participants said could motivate organizational action. Beyond legal concerns, some practitioners reported that being able to use reputational risk as leverage to increase investment in responsible AI work, bringing hypothetical questions like "What if ProPublica found out about ...?" into decision-making meetings. Participant responses in this section illustrate how a reactive organizational stance towards responsible AI work shifts the labor and cost of identifying and addressing issues onto the individual worker.

\subsubsection{Emerging work practices}
In emerging practices on how and when to act, a few  organizations have implemented proactive responsible AI evaluation and review processes for their ML systems, with the work and accountability often distributed across several teams. For example, some respondents reported support and oversight from legal teams. In a few cases, interviewees spoke with enthusiasm about the growing number of both internal and external educational initiatives. This included onboarding and upskilling employees through internal responsible AI curricula to educate employees about responsible AI-related issues and risks as well as externally facing materials to educate consumers and customers. Respondents referred to these efforts as an organization-level proactive investment to set up the organization to better address future responsible AI issues. Furthermore, a few participants shared about the availability or their involvement in preparing externally facing materials to educate their organization's customers or potential customers about responsible AI considerations in practice. A small number of interviewees reported that their work on responsible AI is acknowledged and explicitly part of their compensated role, in contrast to the volunteers in the prevalent practices theme, which is another organization-level difference between prevalent and emerging practices.

On the other hand, emerging practices still show how individuals rather than organizational processes or structures remain the engine of proactive practices. In a few cases, proactive champions organizing grassroots actions and internal advocacy with leadership have made responsible AI a company-wide priority, which then sometimes made it easier for people to get resourcing for responsible AI initiatives and to establish proactive organization-wide processes. Some participants reported leveraging existing internal communication channels to organize responsible AI discussions. One participant even captured screenshots of problematic algorithmic outcomes and circulated them among key internal stakeholders to build support for responsible AI work. Similar to prevalent practices, these individuals are tasked with the labor of using existing organizational structures to build organizational support for their responsible AI work in addition to doing their responsible AI work. The difference in emerging organizational practice is how these individuals are finding more success in instilling a proactive, rather than reactive, mindset for approaching algorithmic responsibility. 

\subsubsection{Mapping the aspirational future}
In an ideal future, many interviewees envisioned organizational frameworks that encourage an anticipatory approach.  In the future state, an individual wanting to engage with algorithmic responsibility issues would not necessarily need to do the  organizational labor of changing structures as in the prevalent and emerging practices, but rather be supported by organization-wide resources and processes to focus their efforts directly on responsible AI work. In this aspirational future, respondents envisioned technical tools to enable large-scale implementation of responsible AI evaluations both internally and externally: well-integrated technical tools would assess algorithmic models developed by product teams and feed seamlessly into organization-wide evaluation processes that identify and address risks of pending ML systems before they go live in products, while externally, customers using the algorithmic models in different contexts have oversight through explicit assessments, which feed information about identified risks back to the organization. Their organizations would utilize clear and transparent communication strategies to explain the process and results of these evaluations both internally within the entire organization and externally with customers and other stakeholders. One practitioner questioned if their team should even engage with customers who do not agree to deploy an assessment framework ex-ante, suggesting a new baseline expectation for customers to also play their role in faster feedback loops for identifying and mitigating risk. Respondents reported that in the ideal future, product managers would have an easier way to understand responsible AI concerns relevant to their products without needing to regularly read large numbers of research papers, which could be supported by organization-level teams, tools, and/or education to synthesize and disseminate relevant knowledge. Several participants expressed that the traditional engineering mindset would need to become better aligned with the dynamic nature of responsible AI issues which cannot be fixed in predefined quantitative metrics. Anticipatory responsible AI frameworks could allow organizations to respond to the responsible AI challenges in ways which uphold organizational code of ethics and society’s values at large.

\subsection{How do we measure success?} \label{measuresuccess}
Another transition we saw our respondents navigating in their work and organizations is how organizations measure success. Many responsible AI initiatives are relatively newer and aim to measure the societal impact of technology, which is a departure from traditional business metrics like revenue or profitability. Learning organizations need to make an active change to better account for this shift. Respondents reported that many challenges in their prevalent work practices arise from the inability to adequately use existing metrics to account for the goals of responsible AI work, while emerging practices aim to begin rewarding success that falls outside of pre-existing narrow definitions. In an aspirational future, organizations value responsible AI work and processes reflect that at every level. 

\subsubsection{Prevalent work practices}
The majority of respondents reported that one of the biggest challenges for their responsible AI work is the lack of metrics that adequately capture its true impact. The majority of respondents also expressed at least some degree of difficulty in communicating the impact of their work. Combined, this hinders them from fully illustrating the importance of responsible AI work for the organization's success, which in turn keeps them from being able to receive adequate credit and compensation for their true impact. The challenges of measuring the impact of responsible AI is a deeply researched topic in the field of fairness, accountability, and transparency of ML. Through our interview questions, we have tried to further disentangle the perspectives on this challenge in industry. For example, some industry practitioners reported that the use of inappropriate and misleading metrics is a bigger threat than the lack of metrics. Respondents shared that academic metrics are very different than industry metrics, which include benchmarks and other key performance indicators tracked by product teams, such as metrics related to customer retention and development (click rate, time spent using a product, etc.) Project managers reported trying to implement academic metrics in order to both leverage academic research and facilitate a collaboration between research and product teams within their organization. One of the interviewees shared that in their personal perspective, "industry-specific product-related problems may not have sufficient research merit or more specifically an ability for the researcher to publish, sometimes because of privacy reasons data used in the research experiments may not allow researchers to be recognized for their work.” This may be due to the nature of the problem or due to privacy reasons. Since data used in the research experiments may not allow researchers to be recognized for their work, this may ultimately discourage them from investigating real world responsible AI issues. Practitioners embedded in product teams explained that they often need to distill what they do into standard metrics such as number of clicks, user acquisition, or churn rate, which may not apply to their work. Most commonly, interviewees reported being measured on delivering work that generates revenue. They spoke at length about the difficulties of measuring responsible AI impact in terms of impact on the business "bottom line.” In some cases, practitioners framed their impact in terms of profitability by arguing that mitigating responsible AI risks prior to launch is much cheaper than waiting for and fixing problems that arise after launch where real-world harm and reputational risk come into play. Again, the prevalent work practices reveal individuals working on responsible AI taking on the extra labor of trying to translate their work into ill-fitting terms and metrics that are not designed to measure or motivate success on responsible AI outcomes.

The majority of respondents expressed at least some degree of difficulty in communicating the impact of their work. The metrics-related challenges they described included: (1) product teams often have short-term development timelines and thus do not consider metrics that aim to encompass long-term outcomes; (2) time pressure within fast-paced development cycles leads individuals to focus on short-term and easier to measure goals; (3) qualitative work is not prioritized because it requires skills that are often not present within engineering teams; (4) leadership teams may have an expectation for "magic," such as finding easy to implement solutions, which in reality may not exist or work; (5) organizations do not measure leadership qualities and (6) do not reward the visionary leaders who proactively address the responsible AI issues that arise; (7) performance evaluation processes do not account for responsible AI work, making it difficult to impossible for practitioners to be rewarded or recognized for their responsible AI contributions.    

\subsubsection{Emerging work practices}
A few interviewees reported that their organizations have implemented metrics frameworks and processes in order to evaluate responsible AI risks in products and services. Practitioners talked enthusiastically about how their organizations have moved beyond ethics washing \cite{bietti2020ethics} in order to accommodate diverse and long-term goals aligned with algorithmic responsibility and harm mitigation, the goals of a responsible AI practice. Interviewees identified the following enablers for this shift in organizational culture: (1) rewarding a broad range of efforts focused on internal education; (2) rewarding risk-taking for the public good; (3) following up on potential issues with internal investigations; (4) creating organizational mechanisms that enable cross-functional collaboration. These emerging organizational enablers begin to set organizational scaffolding of a work environment that supports individuals working on responsible AI as they seek to change how their organization assigns value to work to better align with the societally-focused outcomes. 

\subsubsection{Mapping the aspirational future}
In an aspirational future where responsible AI work is effective and fully supported by organizational structures, interviewees reported that their organizations would measure success very differently than today's prevalent practices: (1) their organizations would have a tangible strategy incorporate responsible AI practices or issues into the key performance indicators of product teams; (2) teams would employ a data-driven approach to manage ethical challenges and ethical decisions in product development; (3) employee performance evaluation processes would be redefined to encompass qualitative work; (4) organizational processes would enable practitioners to collaborate more closely with marginalized communities, while taking into account legal and other socio-technical considerations; (5) what is researched in academic institutions would be more aligned with what is needed in practice; (6) collaboration mechanisms would be broadly utilized. Specifically, participants discussed two kinds of mechanisms to enable collaboration: (1) working with external groups and experts in the field to define benchmarks prior to deployment, and (2) working with external groups to continuously monitor performance from multiple perspectives after deployment.

\subsection{What are the internal structures we rely on?}
In order for individuals to better enable responsible AI work, they need to reexamine the properties of their organizations. This involves leveraging what Orlikowski called the "structurational" model of technology in a specific applied context ~\cite{Orlikowski2000}. In the prevalent practices, organizations do not have internal structures to ensure accountability for responsible AI work, which can then be neglected due to role uncertainty without consequences. Distributed accountability on top of existing structures was reported in emerging practices, while in the aspirational future, responsible AI work would become integrated into all product-related processes to ensure accountability.

\subsubsection{Prevalent work practices}
Most commonly, participants reported ambiguity and uncertainty about role definitions and responsibilities within responsible AI work at their organization, sometimes due to how rapidly the work is evolving. Multiple practitioners expressed that their responsible AI related concerns were heard on account of their seniority in their team and organization. In response to "Do you have autonomy to make impactful decisions?", one data science practitioner who was volunteering time with the responsible AI team shared, "More senior people are making the decisions. I saw ethical concerns but there was difficulty in communicating between my managers and the [responsible AI] team. People weren't open for scrutinization." This illustrates the fragility of the prevalent practice since accountability relies on the individual's own resources, interests, and situational power rather than scalable and systemic organizational structures and processes that would ensure the desired outcomes.
Several interviewees talked about the lack of accountability across different parts of their organization, naming reputational risk as the biggest incentive their leadership sees for responsible AI work, again tying accountability to individual incentives to take responsibility rather than ensuring accountability through organization-wide processes and policies. 
\subsubsection{Emerging work practices}
Interviewees shared these emerging organizational structures as enablers for responsible AI work: (1) flexibility to craft their roles dynamically in response to internal and external factors; (2) distributed accountability across organizational structures and among teams working across the entire product life cycle; (3) accountability integrated into workflows; (4) processes to hold teams accountable for what they committed to; (5) escalation of responsible AI issues to management; (6) responsible AI research groups that contribute to spreading internal awareness of issues and potential solutions; (7) internal review boards that oversee responsible AI topics; (8) publication and release norms that are consistently and widely followed; (9) cross-functional responsible AI roles that work across product groups, are embedded in product groups, and/or collaborate closely with legal or policy teams. Participants also reported being increasingly cognizant of external drivers for change, such as cities and governments participating to create centers of excellence, for example, New York’s Capital District AI Center of Excellence. As before, these emerging structures begin to shift the locus of responsibility for managing organizational change away from the individual who seeks to do responsible AI work, which is not necessarily the same as organizational change management work, and onto organizational processes and structures that can distribute that labor in an appropriate manner.

\subsubsection{Mapping the aspirational future}
 In the future, interviewees envisioned internal organizational structures that would enable responsible AI responsibilities to be integrated throughout all business processes related to the work of product teams. One practitioner suggested that while a product is being developed, there could be a parallel development of product-specific artefacts that assess and mitigate potential responsible AI issues. The majority of interviewees imagined that responsible AI reviews and reports would be required prior to release of new features. New ML operations roles would be created as part of responsible AI audit teams. Currently, this work falls within ML engineering, but respondents identified the need for new organizational structures that would ensure that responsible AI concerns are being addressed while allowing ML engineers to be creative and experiment. For example, one practitioner suggested that a responsible AI operations role could act as a safeguard and ensure that continuous responsible AI assessments are being executed once a system is deployed. Some interviewees described the need for organizational structures that enable external critical scrutiny. Scale could be achieved through partnership-based and multistakeholder frameworks. In the future, public shaming of high-stakes AI failures would provide motivation towards building shared industry benchmarks, and structures would exist to allow organizations to share benchmark data with each other. External or internal stakeholders would need to call out high impact failure use cases to enable industry-wide learning from individual mistakes. Industry-wide standards could be employed to facilitate distributed accountability and sharing of data, guidelines, and best practices. Of note is that in the aspirational future, organizational structures and processes incorporate external parties and perspectives, providing organizations better channels to understand their societal impact.

\subsection{How do we resolve tensions?}
Lastly, responsible AI work brings new types of tensions that organizations may not yet have processes to resolve, especially related to the questions of ethics and unintended consequences of socio-technical systems like AI. This requires organizations to update their prevalent practices in their transitions to better enable responsible AI work. Resolving tensions requires organizations to choose what to prioritize in a situation where there's a need for trade-offs. The practices described below show the different approaches that organizations are taking in prevalent practices, emerging practices, and in the aspirational future.
\subsubsection{Prevalent work practices}
The majority of respondents reported that they see misalignment between individual, team, and organizational level incentives and mission statements within their organization. Often, individuals reported doing ad hoc work based on their own values and personal assessment of relative importance. Similarly, the spread of information relies on individual relationships. Practitioners reported relying on their personal relationships and ability to navigate multiple levels of obscured organizational structures to drive responsible AI work. Related to the question about "What are the ethical tensions that you/your team faces?”, one of the interviewees shared, "We often work on prototypes for specific geographic units which are not meant to be scaled, it’s really meant not to be scaled. We need to step in and make that clear. Also sometimes people state the model is complete, we need a disclaimer that we're still updating and validating it, it is work in progress." Many of the interviewees had to navigate tensions related to scale and expectations on a daily basis. Like in the other transitions, this highlights a prevalent practice of relying on individuals to decide how to resolve tensions rather than organizational processes that would support individuals in evaluating tensions in alignment with the organization's mission or values. This creates additional labor and uncertainty for individuals doing responsible AI work in organizations exhibiting prevalent practices.

\subsubsection{Emerging work practices}
One of the biggest challenges practitioners reported was that as responsible AI ethical tensions are identified, overly rigid organizational incentives may demotivate addressing them, compounded by organizational inertia which sustains those rigid incentives. In this case, although the organizational structures in the emerging work practice shift labor away from individuals onto organization-wide processes, the processes themselves are not sufficiently aligned with the ultimate goals of responsible AI. Therefore, this makes the transition from prevalent to emerging practice one that steers the organization away from, rather than towards, the aspirational future where organizations resolve tensions in a way that encourages responsible AI work. Respondents described that in this situation, research and product teams struggle to justify research agendas related to responsible AI. This was caused by competing priorities that may align better with existing incentives and metrics for success, which as reported in \emph{Section \ref{measuresuccess}: How do we measure success?}, do not adequately account for the impact of responsible AI initiatives.

Interviewees identified several factors that limit an organization's ability to resolve tensions in a manner that enables, instead of hinders, responsible AI work: (1) incentives that reward complexity whether or not it is needed - individuals are rewarded for complex technical solutions; (2) lack of clarity around expectations and internal or external consequences; (3) impact of responsible AI work being perceived as diffuse and hard to identify; (4) lack of adequate support and communication structures - whether interviewees were able to address responsible AI tensions often depended on their network of high trust relationships within the organization; (5) lack of data for sensitive attributes, which can make it impossible to evaluate certain responsible AI concerns. 

\subsubsection{Mapping the aspirational future}
When asked about their vision for the future of their responsible AI initiatives, several respondents wanted responsible AI tensions to be addressed in better alignment with organization-level values and mission statements. They imagined that organizational leadership would understand, support, and engage deeply with responsible AI concerns, which would be contextualized within their organizational context. Responsible AI would be prioritized as part of the high-level organizational mission and then translated into actionable goals down at the individual levels through established processes. Respondents wanted the spread of information to go through well-established channels so that people know where to look and how to share information. With communication and prioritization processes in place, finding a solution or best practice in one team or department would lead to rapid scaling via existing organizational protocols and internal infrastructure for communications, training, and compliance, in contrast to the current prevalent situation that respondents described. Respondents wanted organizational culture to be transformed to enable (1) releasing the fear of being scrutinized as a roadblock for allowing external critical review and (2) distributing accountability for responsible AI concerns across different organizational functions. In the future state, every single person in the organization would understand risk, teams would have a collective understanding of risk, while organizational leadership would talk about risk publicly, admit when failures happen, and take responsibility for broader socioeconomic and socio-cultural implications.

\section{Results: Interdisciplinary Workshop}
As described in the \emph{Section \ref{studyandmethods}: Study and Methods}, after the interviews with practitioners were completed, a workshop was held at a responsible AI oriented venue [anonymized for review]. Each of the four key organizational questions we identified in \emph{Section \ref{results}: Results: Interviews} needs to be considered within the unique socio-technical context of specific teams and organizations - (1) \emph{When and how do we act?} (2) How do we measure success? (3) \emph{What are the internal structures we rely on?} and (4) \emph{How do we resolve tensions?}
However, the literature and interview findings suggest that there are likely similar steps or tactics that could lead to positive outcomes. The workshop activity allowed groups to create landscapes of practices based on their own experiences and then illuminate connections and feedback loops between different practices. Participants were given a simple scenario describing the prevalent work practices and organizational structure of an AI product company in industry as described in \emph{Section \ref{studyandmethods}: Study and Methods}. They then engaged in identifying enablers and tensions elucidating current barriers and pointing the way towards possible solutions. The following themes emerged in the insights participants shared during the activity:

\subsubsection{The importance of being able to veto an AI system}
Multiple groups mentioned that before considering how the fairness or societal implications of an AI system can be addressed, it is crucial to ask whether an AI system is appropriate in the first place. It may not be due to risks of harm, or the problem may not need an AI solution. Crucially, if the answer is negative, then work must stop. They recommended designing a veto power that is available to people and committees across many different levels, from individual employees via whistleblower protections, to internal multidisciplinary oversight committees to external investors and board members. The most important design feature is that the decision to cease further development is respected and cannot be overruled by other considerations.
\subsubsection{The role and balance of internal and external pressure to motivate corporate change}
The different and synergistic roles of internal and external pressure was another theme across multiple groups’ discussions. Internal evaluation processes have more access to information and may provide higher levels of transparency, while external processes can leverage more stakeholders and increase momentum by building coalitions. External groups may be able to apply pressure more freely than internal employees that may worry about repercussions for speaking up.
\subsubsection{Building channels for communication between people (employees and leadership, leadership and board, users and companies, impacted users and companies)}
Fundamentally organizations are groups of people, and creating opportunities for different sets of people to exchange perspectives was another key enabler identified by multiple groups. One group recommended a regular town hall for employees to be able to provide input into organization-wide values in a semi-public forum. 

\subsubsection{Sequencing these actions will not be easy because they are highly interdependent}

Many of the groups identified latent implementation challenges because the discussed organizational enablers work best in tandem. For example, whistleblower protections for employees and a culture that supports their creation would be crucial to ensure that people feel safe speaking candidly at town halls.

It is interesting to observe that workshop discussion groups identified organization-level structures and processes that support and amplify individual efforts as one of the key enablers for responsible AI work. Additionally, these themes are shared as a starting point to spark experimentation. Further pooling of results from trying these recommendations would accelerate learning and progress for all towards achieving positive societal outcomes through scaling responsible AI practices.

\section{Discussion and Conclusion}

As ML systems become more pervasive, there is growing interest and attention in protecting people from harms while also equitably distributing the benefits from these systems. This has led researchers to focus on algorithmic accountability and transparency as intermediary goals on the path to better outcomes. However, corporate responsibility and organizational change are not new themes. The processes we see elucidated by Orlikowski~\cite{Orlikowski2000} and Myserson\cite{meyerson2004tempered} also apply to responsible AI. Myserson described how tempered radicals forge collective action through clarifying issues and creating movements, with a focus on internal culture and actively soliciting support using small but persistent steps. Our interviews and workshop discussions echo similar processes. The results suggest that what individuals working on responsible AI need is for the organizational structures around them to adapt in order to support rather than hinder their work. This can happen as a product of their own advocacy, demonstrated early successes, and/or from leadership proactively steering into these transitions. As Myerson points out, tempered radicals should be aware of new opportunities or threats during their work to elevate social responsibility to an internal corporate priority, and frame their work so it appeals to organizational interests. The resulting tensions in how labor and responsibility are distributed between individuals compared to supporting processes or structures were also prevalent in our findings.

In order to succeed, practitioners have to map out a route from \textit{prevalent work practices} to their \textit{aspirational future state} goals. Along the way, they need to leverage existing practices to build momentum for \textit{emerging work practices} that can lead them there. Similarly, it is essential that practitioners are able to identify and avoid creating \textit{emerging work practices} that work against their desired long-term outcomes. They need to have a clear enough view of what the \textit{aspirational future} should be, while adjusting to changing circumstances. This means both maintaining alignment with the existing organizational state, while maintaining a long-term goal orientation. Our interviews and workshop discussion identified the resulting tensions in getting to that aspirational state. 

Throughout the four key organizational questions in which transitions are necessary to accommodate responsible AI work, we saw that prevalent practices can place the burden of responsibility and labor squarely on individuals to identify issues and try to change outcomes within existing structures.

This means pushing for changes in those structures and processes, as their goals may be antithetical to what is currently supported by the organizational structure. Thus, individuals who want to bring responsible AI issues into their work must do their own jobs, do the responsible AI work if that is not their official job, do the difficult work of redesigning organizational structures around them to accommodate the responsible AI work, and on top of it all, do the change management to get those new organizational practices to be adopted. As a result, incentives may appear misaligned between individuals and their organizational context. This can make it challenging to create adequate support, which should come from communicating the (sometimes small) steps that together make the larger organizational successes. This can invoke feelings of a lack of clarity on expectations and impact. As summarized in \emph{Table \ref{table:1_overview}}, our participants had to decide when and how to act, how to reframe success, orient themselves within internal structures, and resolve tensions between incentives.

Navigating these questions in organizations exhibiting prevalent practices requires skills that are not necessarily part of the regular conversation at academic venues. Perhaps then, rather than focusing on technical complexity, or on calls to (ideal) action alone, we should as a research community also prioritize providing researchers and practitioners with the tools and organizational insight to ensure that they have clear strategies to face this challenge. Researchers who make transitions to industry need to communicate the impact of their work in ways that builds them support within organizations, and legitimacy along the way. They need to have the skills and tools to navigate internal structures and tensions. This requires training, mentorship and sponsorship much beyond technical or research skills. The most effective ways for a particular organization may not always be perfectly aligned with research community norms; perhaps there lies another tension. Rather than having individuals find out the organizational work required on their own, and encounter pitfalls anew as individuals, we can provide support as an insights community, but only when we take this less public work very seriously as a core field of inquiry, as well as education. Perhaps then, this could help us as a wider community to move towards an "open" system, as described in other organizational settings by Scott ~\cite{scott2015organizations}, linking different actors, resources and institutions, and solving complex problems, in similarly complex environments.

We observed that organizations exhibiting the emerging practices were beginning to implement new structures and processes or adapt existing ones, although some emerging structures like rigid organizational incentives within high-inertia contexts can hinder rather than support responsible AI work. The remaining emerging work practices did better enable responsible AI work, often by reducing the labor burden on individuals especially to identify what organizational processes and policies are necessary to support their responsible AI work and to change manage the transitions of adopting those new work practices. This frees up time that individuals can reclaim to focus on the responsible AI work itself.

In the aspirational future, organizational structures and processes would fully provide mechanisms for monitoring and adapting system-level practices to incorporate and address emergent ethical concerns, so individuals who care about algorithmic responsibility issues can easily devote their time and labor to making progress on the specific issues within their functions. The internal advocacy and change management work would be full-time roles given to people with the skills, training, and desire to focus on that work, who could also offer expertise and mentorship to other individuals as they band together to create system-level change inside and beyond their organizations. Individuals working on responsible AI would then be free to focus on their specific job, rather than on changing the job environment to make it possible to do their job.

The impact of ML systems on people cannot be changed without considering the people who build them and the organizational structure and culture of the human systems within which they operate. A qualitative methodological approach has allowed us to build rich context around the people and organizations building and deploying ML technology in industry. We have utilized this qualitative approach here in order to investigate the organizational tensions that practitioners need to navigate in practice. We describe existing enablers and barriers for the uptake of responsible AI practices and map a transition towards an aspirational future that practitioners describe for their work. In line with earlier organizational  research, we emphasize such transitions are not to be seen as linear movements from one fixed state to another, rather they represent persistent steps and coalition building within the ever-changing nature of organizational contexts themselves.

\bibliographystyle{ACM-Reference-Format}
\bibliography{references.bib}

\received{June 2020}
\received[revised]{October 2020}
\received[accepted]{December 2020}

\appendix
\section{Questionnaire} \label{questionnaire}
\subsection{Describing current work practices related to fairness, accountability, and transparency of ML products and services}
\subsubsection{Describe your role}
\begin{enumerate}
    \item What is your formal role by title?
    \item How would you describe your role?
    \item Is your formal role matched to your actual role? 
        \begin{itemize}
        \item If not, how is it not?
        \end{itemize}
    \item Is your organization flexible in the way it sees roles?
        \begin{itemize}
        \item If not, what is it like?
        \end{itemize}
    \item How did you assume your role? 
        \begin{itemize}        
        \item If hired in, were you hired externally or transitioned?
        \item If you transitioned, where did you transition from?
        \item Does your company generally move people around fluidly?
        \item Does your company reward broad knowledge/skills across different industries or specializations?
        \end{itemize}
    \item How did your role change over time?
        \begin{itemize}        
        \item From a responsibility perspective?
        \item From a people perspective?
        \end{itemize}
    \item Is role scope change typical at your company?
    \item If yes, what does it typically look like - is it…
        \begin{itemize}
        \item Scope creep? 
        \item Is it explicitly within your job description?
        \item Planned role expansion?
        \item Stretch assignments?
        \end{itemize}
    \item Do you have autonomy to make impactful decisions?
        \begin{itemize}
        \item If yes, how?
        \item If no, what is the case instead?
        \end{itemize}
\end{enumerate}

\subsubsection{How did your fairML effort start?}
\begin{enumerate}
    \item Was it officially sponsored?
        \begin{itemize}
            \item If yes, by whom - what level of leader?
            \item If no, who launched the effort - was it a team? A motivated individual? What level of leadership?
        \end{itemize}
    \item Why did the effort start?
    \item Was the effort communicated to employees?
        \begin{itemize}
            \item Who was it communicated to?
            \item How was it communicated?
        \end{itemize}
    \item Is the effort part of a program or stand-alone? 
    \item Is it tied to a specific product’s development or launch?
        \begin{itemize}
            \item What is the product?
            \item What is its primary use case?
            \item Who is the primary end user?
            \item When is it slated to launch?
        \end{itemize}
    \item Are you part of a team or doing this kind of work by yourself?
    \item Is it a volunteering effort?
        \begin{itemize}
            \item If so, are you getting rewarded or recognized for your time? How?
        \end{itemize}
    \item What types of activities have been done or are planned?
    \item Are you actively collaborating with external groups? What groups and why?
\end{enumerate}

\subsubsection{Responsibility and accountability}
\begin{enumerate}
    \item Who is accountable for aspects around risk or unintended consequence…
    \begin{itemize}
        \item Identifying risk?
        \item Solutioning against risk?
        \item Fixing mistakes?
    \end{itemize}
    \item Avoiding negative impact, including press? Is your sponsor connected to risk management? 
    \begin{itemize}
        \item How so?
        \item What is their level of accountability relative to risk?
        \item What are they responsible for doing?
    \end{itemize}
    \item Who are your main stakeholders? 
    \item What are the other departments you work with? 
        \begin{itemize}
        \item How has that changed since the effort launched? 
        \item What was the business case for the fairML work/team?
        \item What teams are adjacent to this effort (ie, not directly involved but “friends and family”) - is there an active compliance function, eg?
        \item (if no answer, probe for e.g. product teams, compliance, trust and safety type teams, or value-based design efforts, ethics grassroots etc)
        \item What are other efforts in your organization that are similar to Accountability work for instance Diversity \& Inclusion and what does that look like? Is there general support for this type of effort?
        \end{itemize}
        \item Do you feel there is support for this effort?
        \begin{itemize}
            \item Why or why not?
            \item Who supports it (what company career level, function, role and/or geography)?
            \item Who doesn’t support it?
        \end{itemize}
        \item Would you say this effort aligns to company culture? How or how not?
        \item Is scaling possible?
        \begin{itemize}
            \item If so, do you intend to scale?
            \item If not, why not?
        \end{itemize}
\end{enumerate}

\subsubsection{Performance, rewards, and incentives}
\begin{enumerate}
    \item How is performance for your Algorithmic Accountability effort defined at your company?
    \item What are you evaluated on in your role?
    \item What works about the way performance is measured? What are some flaws?
    \item What does your performance management system/compensation structure seek to incentivize people to do (what is the logic behind the approach)?
    \item What does your performance management system/compensation structure actually incentivize people to do?
    \item What kind of person gets consistently rewarded and incentivized?
\end{enumerate}

\subsubsection{Risk culture}
\begin{enumerate}
\item How do you work with external communication teams - PR, Policy?
\begin{itemize}
    \item Who owns that relationship - is it a centralized team?
    \item What is that comms teams primary accountability (eg, press releases, think pieces, etc)?
    \item Has the team managed risk before?
    \item Is the team mobilized to manage risk?
\end{itemize}

\item How do you work with Legal?
\begin{itemize}
    \item Is it a visible function in the organization?
    \item Does it have authority to make decisions and company policy, from your PoV?
\end{itemize}
\begin{itemize}
    \item How do you engage with communities? 
    \item What types of communities?
    \item What does this look like?
    \item What types of communication have you set up?
\end{itemize}
\item What are the ethical tensions that you/your team faces? 
\item On a scale of 1-5, what is your level of perception of your company’s risk tolerance?
\end{enumerate}

\subsection{Future dream state - a structured way of getting a mapping of future dream state}
\begin{enumerate}
    \item What is your company’s current state for fairML practice? (people, process, technology)
    \item What is your vision for the future state of the fairML practices? 
    \item What do you need to change to get to the future state? 
    \item What do you need to retire to get to the future state?
    \item What can be salvaged/repurposed?
\end{enumerate}

\subsection{Ending notes}

\begin{enumerate}
    \item What is the best about your current set up?
    \item How would you summarize the largest challenges? Aka what do you like least?
    \item Is there anything that I should have asked about?
\end{enumerate}

\end{document}